\def\a{\alpha}\def\c{\chi}\def\e{\epsilon}

\def
\p{\pi}\def\q{\psi}\def\t{\tau}

\def\G{\Gamma}
\def\O{\Omega}\def\S{\Sigma}

\def\per{\times}
\def\mo{{-1}}\def\ha{{1\over 2}}
\def\qu{{1\over 4}}

\def\ds{ds^2=}

\def\af{asymptotically flat }\def\st{spacetime }
\def\fe{field equations }\def\bh{black hole }\def\as{asymptotically }
                                   
\def\bg{background }\def\gs{ground state }

\def\cco{cosmological constant }\def\em{electromagnetic }
\def\ssy{spherically symmetric }
\def\ms{maximally symmetric }

\def\dof{degrees of freedom }

\def\ads{anti-de Sitter }

\def\des{de Sitter }
 
\def\GB{Gauss-Bonnet }

\def\section#1{\bigskip\noindent{\bf#1}\smallskip}
\def\nota{\footnote{$^\dagger$}}

\def\PL#1{Phys.\ Lett.\ {\bf#1}} 
\def\PRL#1{Phys.\ Rev.\ Lett.\ {\bf#1}} 
\def\PR#1{Phys.\ Rev.\ {\bf#1}}\def\CQG#1{Class.\ Quantum Grav.\ {\bf#1}} 
\def\NP#1{Nucl.\ Phys.\ {\bf#1}} 
\def\JMP#1{J.\ Math.\ Phys.\ {\bf#1}}

\def\ref#1{\medskip\everypar={\hangindent 2\parindent}#1}
\def\beginref{\begingroup
\bigskip
\centerline{\bf References}
\nobreak\noindent}
\def\endref{\par\endgroup}

\def\dc{{1\over16\p}\int d^5x\, \sqrt{-g}\ }\def\ab{\bar \a}
\def\ec{-{1\over16\p}\int d^5x\, \sqrt g\ }
\def\IN{I_{S^2\times S^3}}\def\ID{I_{S^5}}\def\tm{{3/2}}

\magnification=1200\baselineskip18pt

{\nopagenumbers
\line{June 1998\hfil INFNCA-TH9806}
\vskip40pt
\centerline{\bf Charged gravitational instantons in five-dimensional}
\smallskip
\centerline{\bf Einstein-Gauss-Bonnet-Maxwell theory}
\vskip40pt
\centerline{\bf S. Mignemi}
\vskip10pt
\centerline{Dipartimento di Matematica, Universit\`a di Cagliari}
\centerline{viale Merello 92, 09123 Cagliari, Italy}
\centerline{and INFN, Sezione di Cagliari}
\vskip80pt
{\noindent
We study a solution of the Einstein-\GB theory in 5 dimensions coupled
to a Maxwell field, whose euclidean continuation gives rise to an instanton
describing black hole pair production. We also discuss the dual theory
with a 3-form field coupled to gravity.
}

\vfil\eject}

\section{1. Introduction.}
Vacuum solutions of four-dimensional Einstein gravity with positive \cco
are characterized by the presence of a cosmological horizon.
In case of vanishing mass the solutions are given by the maximally symmetric
\des\st, but for nonzero mass one can obtain \ssy \bh solutions which contain
an event horizon besides the cosmological one.
In the limiting case in which the two horizons coincide these solutions
reduce to a non-trivial spacetime metric, known as Nariai metric, which has
topology $H^2\per S^2$. Its euclidean continuation describes a gravitational 
instanton which can be used to compute the creation rate for black hole
pairs in a cosmological background [1]. When a Maxwell field is coupled, more
general black hole solutions can also be obtained carrying electric or
magnetic charge.
In this case one can have up to three horizons, and a few special
solutions giving rise to instantons by euclidean continuation can be found
[2-3].

In higher dimensions, the Einstein-Hilbert action can be generalized by 
the addition of the so called Gauss-Bonnet terms [4-5]. The generalized 
action gives rise to field
equations which are still second order and no new \dof are introduced in the 
theory. In four dimensions or less the \GB terms are total derivatives and do
not contribute to the field equations.
It is well known that the actions so generalized admit \as\des
solutions even in absence of a cosmological constant.
In particular, the most general \ssy solutions of the Einstein-\GB theory
coupled to a Maxwell field
have been obtained in arbitrary dimensions [6]. They include solutions
containing both cosmological and event horizons. One may thus
wonder if solutions with properties similar to the Nariai metric
or its charged generalizations are avalaible also in this case for special
values of the parameters.

In this paper we consider the simplest non-trivial example, namely the
five-dimensional theory. This case is quite peculiar, since contrary to
the higher-dimensional ones, the metric function has only one root,
leading in general to the appearence of naked singularities, and so it turns
out that the only physically interesting case is that in which the root is
double, corresponding to two coinciding horizons. The properties of this
metric are therefore similar to those of the Nariai solution.
On the other hand, five dimensions is the only case that can be 
treated analitically, because in higher dimensions
one cannot in general find a closed algebraic expression for the
location of the horizons.

\section {2. The lorentzian solution.}
The five dimensional Einstein-\GB-Maxwell action is 
$$I=\dc (R+\ab S-\qu F^2), \eqno(1)$$
where $S=R_{abcd}R^{abcd}-4R_{ab}R^{ab}+R^2$ is the \GB term and $\ab$ is a
coupling constant.
The \fe are\nota{Throughout this paper, we adopt orthonormal indices.}
$$G_{ab}+\ab S_{ab}=T^F_{ab},\eqno(2)$$
where 
$$\eqalignno{S_{ab}&=2R_{acde}R_b^{\ cde}-4R_{acbd}R^{cd}-4R_{ac}R_b^{\ c}
+2RR_{ab},&(3)\cr
T^F_{ab}&=\ha F_{ac}F^c_{\ b}-{1\over8}F^2g_{ab}.&(4)\cr}$$
The theory admits two \ms solutions with vanishing charge, namely flat space 
and 5-dimensional \des or \ads space (depending on the sign of $\ab$),
with curvature $R=-10/\ab$.
These can be considered as the ground states of the theory. 
It has been argued [5] that the \des \bg is unstable under small 
perturbations, so that the actual \gs is flat Minkowski space. However,
in this paper we are mainly interested in the \des sector of the theory.
Hence, we shall consider the case $\ab<0$, which admits the existence 
of a cosmological horizon, and ignore the \af solution.
Charged \ssy solutions are given by [6]
$$F_{01}={Q\over r^3},\eqno(5)$$
$$ds^2=-V(r)dt^2+V^\mo(r)dr^2+r^2d\O^2_3,\eqno(6)$$
with
$$V(r)=1-{r^2\over2\a}\left[1\pm\sqrt{1-4\a\left({2m\over r^4}-{q^2\over r^6}
\right)}\right],\eqno(7)$$
where $q^2=Q^2/12$ and $\a=-2\ab>0$.
The branch with the minus sign is \af, while the branch with the plus sign,
which we shall consider in the following, is \as \des. In the last case, a
horizon can be present if $-\a<2m<3\a$. More precisely, if $4q^2>(\a+2m)^2$,
$V(r)$ is always negative. If instead $4q^2\le(\a+2m)^2$, $V(r)$ has a unique 
zero at $r_0^2={\a+2m-\sqrt{(\a+2m)^2-4q^2}\over2}$. In this case, however, 
a branch singuarity is present at a point $r_s>r_0$, where the square root 
in (7) becomes negative. The only possibility to avoid this singularity is
in the extremal case, $4q^2=(\a+2m)^2$, when $r_0$ is a double root.
In this case apparently there is no physical region, since one has two
coinciding horizons. However, this is not the case, as can be shown by
adopting
a limit procedure similar to that originally discussed in [1] for the Nariai
solution.

In fact, for $4q^2\to(\a+2m)^2$, one can expand $V(r)$ in a neighborood of
$r_0=\sqrt{\a+2m\over2}$ as
$$r=r_0\left(1+{\e\over2}\cos\c+o(\e^2)\right),\eqno(8)$$
so that 
$$r^2={\a+2m\over2}\left(1+\e\cos\c+o(\e^2)\right),\eqno(9)$$
where $\c$ is a new coordinate.
One can also expand $q^2$ as $q^2={(\a+2m)^2\over4}(1-\e^2)$ and
define a rescaled time coordinate
$$\t={8\sqrt{\a+2m}\over3\a-2m}\e t.\eqno(10)$$
Substituting in (6), one finally gets, at leading order in $\e$,
$$\ds{1\over A}(-\sin^2\c d\t^2+d\c^2)+{1\over B}d\O^2_3,\eqno(11)$$
where
$$A={8\over3\a-2m}={4\over2\a-|q|},\qquad\qquad B={2\over\a+2m}={1\over|q|},
\eqno(12)$$
Moreover, in this limit, $F_{01}=QB^{3/2}=2\sqrt{12}/(\a+2m)$.
The constants $A$ and $B$ are both positive since, as discussed previously,
$-\a<2m<3\a$ (or equivalently $|q|<2\a$).
The metric (11) has the form of a product of a two-dimensional \des space of
size $1/\sqrt A$ with a three-sphere of radius $1/\sqrt B$ and is therefore 
perfectly regular. Its Penrose diagram is that of two-dimensional \des
spacetime. In particular, in analogy with the Nariai metric [1], an observer
sees two cosmological horizons,
both in the positive and negative $\c$-directions. The solution can therefore
be interpreted as a pair of black holes at antipodal points
on the spatial section of a \des \st. 

The previous solution can also be obtained directly from the field equations
(2), which for a metric of the form (11) and electric field $F_{01}=QB^{3/2}$
reduce to
$$-3B=-{Q^2\over4}B^3,\qquad\qquad-A(1-2\a B)-B={Q^2\over4}B^3,\eqno(13)$$
and hence, recalling that $Q^2=12q^2$, $B={1\over|q|}$, $A={4\over2\a-|q|}$,
in accordance with (12).

From (13) follows that in five dimensions a non-flat solution
is avalaible only if $Q\ne0$.

The metric (11) can also be obtained by duality in the case of Einstein-\GB
gravity coupled to a 3-form field $H_{abc}$, with action
$$I=\dc (R+\ab S-{1\over12}H^2).\eqno(14)$$
The gravitational \fe are now
$$G_{ab}+\ab S_{ab}=T^H_{ab},\eqno(15)$$
with
$$T^H_{ab}=\qu H_{acd}H_b^{\ cd}-{1\over24}H^2g_{ab}.\eqno(16)$$
With the ansatz $H_{abc}=QB^\tm\e_{abc}$, one has
$T^H_{ab}=T^F_{ab}$, and hence the \fe reduce to those of the \em case.

\section{3. The euclidean metric.}
The euclidean continuation of the metric (11) is given by
$$\ds{1\over A}(\sin^2\c d\q^2+d\c^2)+{1\over B}d\O^2_3,\eqno(17)$$
where $\q=i\t$ is the euclidean time and $0\le\q<2\p$, $0\le\c<\p$.
The metric (17) describes the product of two round spheres,
$S^2\times S^3$. In analogy with the charged Nariai metric in four
dimensions [1], it can be interpreted as a gravitational instanton mediating
the creation of a pair of black holes in a \bg\des\st.

As discussed in ref. [3], according to the no-boundary proposal,
the pair creation rate can be estimated as
$$\G=\exp[-2(\IN-\ID)],\eqno(18)$$
where $\IN$ is the action of the half-instanton with boundary the spacelike
surface $\S$ of topology $S^2\per S^2$, corresponding to the maximal spatial
section of (11), and $\ID$ is the action of the half euclidean \des space
of radius $\sqrt\a$, with spatial boundary $S^4$.

In the case of the Maxwell field, the euclidean action is given by
$$I=\ec(R+\ab S-\qu F^2)+{\rm boundary\ terms}.\eqno(19)$$
The gravitational boundary terms do not contribute, since in the case
under study the second fundamental form vanishes on the boundaries, 
but an electric contribution should be added in order to keep the charge 
fixed at the boundary [3]. This has the form 
$$I_b={1\over16\p}\int_\S d^4x\ \sqrt h\ F^{ab}n_a A_b,\eqno(20)$$
where $n^a$ is the outgoing normal to $\S$.
Using the trace of the \fe, the five-dimensional integral in (19) reduces to
$$I={1\over8\p}\int d^5x\ \sqrt g\ R .\eqno(21)$$
For $S^5$, eq. (21) gives $I={5\over4\p\a}Vol(S^5)={5\p^2\over8}\a^\tm$, 
where $Vol(\ )$ is the volume of the half-instanton. It is important to 
notice that, due to the contribution of the \GB term, the sign of $I$ is
opposite to the one that would have been obtained in the case of Einstein
theory with a cosmological constant. The high value of the action is 
another sign of the instability of the \des space in the Einstein-\GB
model.
For $S^2\per S^3$, $I={A+3B\over8\p\a}Vol(S^2\per S^3)=
\p^2{A+3B\over AB^\tm}$, 
and moreover $I_b$ gives a further contribution ${3\p^2\over AB^\tm}$. 
The total action is therefore
$$\IN=\p^2{A+6B\over AB^\tm}={\p^2\over2}\sqrt{|q|}(6\a-|q|).\eqno(22)$$
The value of the action goes to zero for $q\to0$, since the volume of the
instanton vanishes in this limit.
Finally, substituting in (18),
$$\G=\exp\left[-\p^2\left(\sqrt{|q|}(6\a-|q|)-{5\over4}\a^\tm\right)\right].
\eqno(23)$$
In the interval of allowed values for $|q|$, $0<|q|<2\a$, the exponent of
$\G$ can assume
both positive and negative sign. In particular, it is positive for small
$|q|$, corresponding to a high production rate. Five dimensional \des space 
appears therefore to be unstable in
our model, for creation of black hole pairs of charge $|q|\ll\a$.

The same calculation can be done for the 3-form field $H$. In this case,
the euclidean action is
$$I=\ec (R+\ab S-{1\over12}H^2)= {1\over8\p}\int d^5x\ \sqrt g\ (R+
{1\over12}H^2),\eqno(24)$$
where the trace of the \fe has been used. Also in this case the gravitational
boundary terms vanish, while no boundary term is necessary for the 3-form
field [3]. Substituting the values of the fields in (24), one gets the same
result (22) obtained in the Maxwell case. It appears therefore that one can
extend also to higher dimensions the validity of the conjectures of [3]
on the invariance under duality of the pair production rate of black holes.

\section{4. Conclusion.}
We conclude with some considerations on the higher-dimensional generalizations
of the results discussed here. Although in principle straightforward, the
possibility of making
explicit calculations is prevented because one can no longer obtain an
explicit expression for the location of the horizon, since one should solve
algebraic equations of higher degree.
One can show however an important qualitative difference in higher
dimensions, since in $D\ge6$ it is possible to get two distinct horizons and
hence a
physical region even in the non-extremal case. Thus solutions with properties
similar to the lukewarm black holes of ref. [2] are possible.

One can still, however, look directly for solutions of the $D$-dimensional
theory of the form deS$^2\per S^d$, with $d=D-2$,
which generalize the one we have discussed in this paper.
In this case, the gravitational \fe for a \em field $F_{01}=Q$
or a $d$-form field $F_{abc...}=Q\e_{abc...}$ give:
$$\eqalign{-&\ha d(d-1)B[1+(d-2)(d-3)\ab B]=-{Q^2\over4},\cr
-&A[1+2(d-1)(d-2)\ab B]-\ha(d-1)(d-2)B[1+(d-3)(d-4)\ab B]={Q^2\over4}.\cr}
\eqno(25)$$

For $d>3$ (i.e. $D>5$), one has solutions even if $Q=0$, with
$B=-[(d-2)(d-3)\ab]^\mo$, $A=-(d-1)[(d+1)(d-2)\ab]^\mo$,
while, in general, a 1-parameter class of solutions is available for $Q\ne0$.
This also includes the special case $A=B$. It remains an open question if these
solutions can still be interpreted as limiting cases of \bh solutions.
\bigskip
\beginref
\ref[1] P. Ginsparg, M.J. Perry, \NP{B222}, 245 (1983);
\ref[2] F. Mellor and I. Moss, \PL{B222}, 361 (1989); \CQG{6}, 1379 (1989);
L.J. Romans, \NP{B383}, 395 (1992);
\ref[3] S.W. Hawking and S.F. Ross, \PR{D52}, 5865 (1995);
R.B. Mann and S.F. Ross, \PR{D52}, 2254 (1995);
\ref[4] D. Lovelock, \JMP{12}, 498 (1971);
\ref[5] D.G. Boulware and S. Deser, \PRL{55}, 2656 (1985);
\ref[6] D.L. Wiltshire, \PL{B169}, 36 (1986); \PR{D38}, 2445 (1988).
\endref
\end